# Broadband Macroscopic Cortical Oscillations Emerge from Intrinsic Neuronal Response Failures

**Amir Goldental[1,†], Roni Vardi[2,†], Shira Sardi[1,2], Pinhas Sabo[1] and Ido Kanter[1,2,*]**

[1]Department of Physics, Bar-Ilan University, Ramat-Gan 52900, Israel

[2]Gonda Interdisciplinary Brain Research Center and the Goodman Faculty of Life Sciences, Bar-Ilan University, Ramat-Gan, Israel

†These authors contributed equally to this work.

*Correspondence: ido.kanter@biu.ac.il

**Abstract**

**Broadband spontaneous macroscopic neural oscillations are rhythmic cortical firing which were extensively examined during the last century, however, their possible origination is still controversial. In this work we show how macroscopic oscillations emerge in solely excitatory random networks and without topological constraints. We experimentally and theoretically show that these oscillations stem from the counterintuitive underlying mechanism - the intrinsic stochastic neuronal response failures. These neuronal response failures, which are characterized by short-term memory, lead to cooperation among neurons, resulting in sub- or several- Hertz macroscopic oscillations which coexist with high frequency gamma oscillations. A quantitative interplay between the statistical network properties and the emerging oscillations is supported by simulations of large networks based on single-neuron *in-vitro* experiments and a Langevin equation describing the network dynamics. Results call for the examination of these oscillations in the presence of inhibition and external drives.**

## 1. INTRODUCTION

The most widespread cooperative activity of neurons within the cortex is spontaneous macroscopic oscillations(Silva et al., 1991;Gray, 1994;Contreras et al., 1997;Buzsaki and Draguhn, 2004;Chialvo, 2010), which range between sub- and hundred- Hertz(Başar et al., 2001;Brovelli et al., 2004;Buzsaki and Draguhn, 2004;Grillner et al., 2005;Giraud and Poeppel, 2012). The high cognitive functionalities of these oscillations are still controversial(Klimesch, 1996;1999;Başar et al., 2001;Wiest and Nicolelis, 2003;Buzsaki, 2006;Kahana, 2006;Fries, 2009;Giraud and Poeppel, 2012;Thivierge et al., 2014) and are typically attributed to transitory binding activities among indirect macroscopic distant cortical regions(Gray, 1994;Başar et al., 2001;Buzsaki and Draguhn, 2004;Roxin et al., 2004;Fries, 2009). In addition, it was found that the theta rhythms(Klimesch, 1999;Buzsaki and Draguhn, 2004), oscillations in the range of 4-10 Hz, play a key role in the formation and retrieval of episodic and spatial memory(Hasselmo, 2005). This theta rhythm is usually accompanied by high frequency oscillations in the range of 30-80 Hz, known as gamma oscillations(Colgin and Moser, 2010). Gamma oscillations are also related to sensory stimulations and induce neuronal ensemble synchrony by generating a narrow window for effective excitation(Cardin et al., 2009).

There are several suggested mechanisms for the formation of such rhythms on the network level(Wang, 2010). Most of the proposed mechanisms are based on the existence of inhibitory synapses(Wilson and Cowan, 1972;Jirsa and Haken, 1996;Brunel and Wang, 2003), especially for high frequency oscillations(Brunel and Wang, 2003;Colgin and Moser, 2010;Wang, 2010). For illustration, assume that a fast excitation increases neural firing in an excitatory short-delayed feedback loop. Consequently, neuronal populations along the excitatory feedback loop will fire at high rates that will cause a slower response of the inhibitory neurons. As a result, the inhibitory neurons will depress the activity within the excitatory population. This will then depress the excitation of the inhibitory population. Finally, the depression of the inhibitory neurons allows a repeated fast excitation of the excitatory population.

In this work we show how extra-cellular potential oscillations, synchronized rhythmic firing of neurons, emerge in random networks without inhibitory synapses. Our findings are based on an experimental observations of neuronal plasticity in the form of intrinsic neuronal response failures(Vardi et al., 2015). Using simulations of large networks, based on single-neuron *in-vitro* experiments, we show that this type of neuronal plasticity leads to the coexistence of both theta and gamma oscillations. Results are supported by a quantitative approach based on a Langevin equation, which describes the network dynamics.

## 2. MATERIALS AND METHODS

### 2.1 *In-vitro* experiments

#### 2.1.1 Animals

All procedures were conducted in accordance with the National Institutes of Health Guide for the Care and Use of Laboratory Animals and Bar-Ilan University Guidelines for the Use and Care of Laboratory Animals in Research and were approved and supervised by the Institutional Animal Care and Use Committee.

#### 2.1.2 Culture preparation

Cortical neurons were obtained from newborn rats (Sprague-Dawley) within 48 h after birth using mechanical and enzymatic procedures. The cortical tissue was digested enzymatically with 0.05% trypsin solution in phosphate-buffered saline (Dulbecco's PBS) free of calcium and magnesium, and supplemented with 20 mM glucose, at 37°C. Enzyme treatment was terminated using heat-inactivated horse serum, and cells were then mechanically dissociated. The neurons were plated directly onto substrate-integrated multi-electrode arrays (MEAs) and allowed to develop functionally and structurally mature networks over a time period of 2-3 weeks *in-vitro*, prior to the experiments. Variability in the number of cultured days in this range had no effect on the observed results. The number of plated neurons in a typical network was in the order of 1,300,000, covering an area of about 380 mm$^2$. The preparations were bathed in minimal essential medium (MEM-Earle, Earle's Salt Base without L-Glutamine) supplemented with heat-inactivated horse serum (5%), glutamine (0.5 mM), glucose (20 mM), and gentamicin (10 g/ml), and maintained in an atmosphere of 37℃, 5% $CO_2$ and 95% air in an incubator as well as during the electrophysiological measurements.

### 2.1.3 Synaptic blockers

All experiments were conducted on cultured cortical neurons that were functionally isolated from their network by a pharmacological block of glutamatergic and GABAergic synapses. For each culture 20 μl of a cocktail of synaptic blockers was used, consisting of 10 μM CNQX (6-cyano-7-nitroquinoxaline-2,3-dione), 80 μM APV (amino-5-phosphonovaleric acid) and 5 μM bicuculline. This cocktail did not block the spontaneous network activity completely, but rather made it sparse. At least one hour was allowed for stabilization of the effect.

### 2.1.4 Stimulation and recording

An array of 60 Ti/Au/TiN extracellular electrodes, 30 μm in diameter, and spaced 500 μm from each other (Multi-Channel Systems, Reutlingen, Germany) was used. The insulation layer (silicon nitride) was pre-treated with polyethyleneimine (0.01% in 0.1 M Borate buffer solution). A commercial setup (MEA2100-2x60-headstage, MEA2100-interface board, MCS, Reutlingen, Germany) for recording and analyzing data from two 60-electrode MEAs was used, with integrated data acquisition from 120 MEA electrodes and 8 additional analog channels, integrated filter amplifier and 3-channel current or voltage stimulus generator (for each 60 electrode array). Mono-phasic square voltage pulses typically in the range of [-800, -500] mV and [60, 400] μs were applied through extracellular electrodes. Each channel was sampled at a frequency of 50k samples/s, thus the changes in the neuronal response latency were measured at a resolution of 20 μs.

### 2.1.5 Cell selection

Each node was represented by a stimulation source (source electrode) and a target for the stimulation – the recording electrode (target electrode). These electrodes (source and target) were selected as the ones that evoked well-isolated, well-formed spikes and reliable response with a high signal-to-noise ratio. This examination was done with a stimulus intensity of -800 mV with duration of 200 μs using 30 repetitions at a rate of 5 Hz, followed by 1200 repetitions at a rate of 10 Hz.

### 2.1.6 *In-vitro* experiment with feedback loops and neural circuits

The activity of all source and target electrodes was collected and action potentials were detected on-line by threshold crossing, and entailed stimuli were delivered in accordance with the circuit's connectivity, as described below. A successful response was defined as a spike occurring within a typical time window of 2-10 ms following the beginning of an electrical stimulation.

In **Figures 2A,B**, after every spike detection two supra-threshold extracellular stimulations were given to the same neuron, after 600 ms and 630 ms. In case that the timings of the stimulations overlap, only one stimulation is given.

In **Figures 2C,D**, after every spike detection supra-threshold extracellular stimulations were given to its connected neurons. For example, if a spike was detected at the left (green) neuron (**Figures 2C**), a supra-threshold extracellular stimulation will be given to the middle (brown) neuron after 330 ms.

### 2.1.7 Data analysis

Analyses were performed in a Matlab environment (MathWorks, Natwick, MA, USA). The reported results were confirmed based on at least eight experiments each, using different sets of neurons and several tissue cultures.

The temporal firing frequency, around stimulation no. i, of the neuron in Figure 1 was calculated using the following procedure

$$firing\ frequency\ (i) = stimulation\ frequency \cdot \sum_{m=\max(0, i-125)}^{i+125} \frac{Is\_Spiked(m)}{i + 125 - \max(0, i - 125)}$$

where Is_Spiked(m) = 1 if the neuron responded to stimulation no. m, otherwise Is_Spiked(m) = 0.

### 2.2 Simulations

Simulations (similar to (Vardi et al., 2015)) consist of a network of N leaky integrate and fire neurons

$$\frac{dV_i}{dt} = -\frac{V_i}{\tau} + \sum_{j=1}^{N} J_{ji} \sum_{t'} \delta(t - t' - D_{ji})$$

where $i \in [1,N]$, $\tau$=20 ms, $J_{ji}$ and $D_{ji}$ are the connection's strength and delay from neuron j to i, respectively. The summation over t' sums all firing times of neuron j, the integration time step is 0.05 ms, and the threshold is 1. For the $n^{th}$ threshold crossing of a neuron, its probability for a response is $[\Sigma(\tau_{n-m}/\tau_c)\exp(-\alpha m)]/[\Sigma\exp(-\alpha m)]$, where $\tau_n$ is the time gap between the $n^{th}$ and $(n-1)^{th}$ threshold crossings, $\tau_c$=1/$f_c$, $\alpha$=1.4 and the sum is over m≥0. A refractory period of 2 ms is imposed after an evoked spike, for response failures the voltage is set to 0.2. Results were found to be insensitive to initial conditions.

### 2.2.1 Various forms of p(s|τ) lead to the same <ISI>

The probability for a response, given the last inter-stimulation-interval, p(s|τ), should lead to <ISI>=$\tau_c$ (**Figures 1** and **3**). One can show that any p(s|τ) satisfying

$$\frac{\int_0^\infty p(s|\tau)p(\tau)d\tau}{\int_0^\infty \tau p(\tau)d\tau} = \frac{1}{\tau_c}$$

where p(τ) is the probability density of an inter-stimulation-interval equals to τ, leads to <ISI>=τ$_c$. The numerator on the left hand side stands for the average probability for a successful response, and the denominator stands for the average inter-stimulation-interval. This ratio is equivalent to the firing rate, hence equals to 1/τ$_c$. For instance p(s|τ)=τ/τ$_c$, this theoretical curve fits all p(τ) (**Figure 3D**). In the activity of some random networks a good approximation for p(τ) is

$$p(\tau) = \frac{2}{\tau_c} exp\left(-\frac{2\tau}{\tau_c}\right).$$

For this p(τ) some of the p(s|τ) solutions, which lead to <ISI>=τ$_c$, are p(s|τ)=0.5, p(s|τ)=τ/τ$_c$ and

$$p(s|\tau) = 1 - exp\left(-\frac{2\tau}{\tau_c}\right),$$

which is similar to **Figure 3D**.

### 2.2.2 Fourier analysis of the rate

To perform a Fourier analysis on the activity of the network we define the rate vector:

$$R(time = i \cdot w) = \frac{1}{Nw} \int_{(i-0.5)w}^{(i+0.5)w} \sum_{t_{spike}} \delta\left(t' - t_{spike}\right) dt'$$

where i is an integer, w is a predefined time window and the sum is over all spike times of all N neurons. Next, a discrete Fourier transform is preformed and the function of resulting amplitudes is normalized and smoothed using a sliding window of 1 Hz.

## 3. RESULTS

### 3.1 Neuronal plasticity: Neuronal response failures

We start with the quantification of the neuronal response latency (NRL)(Vardi et al., 2013a;Marmari et al., 2014;Vardi et al., 2015), measured as the time-lag between a stimulation and its corresponding evoked spike (**Figure 1**). It was recently shown(Vardi et al., 2015) that when a neuron is repeatedly stimulated, its NRL stretches gradually (**Figure 1**, upper panel), and when the stimulation frequency is high enough stochastic neuronal response failure (NRFs) emerge. Specifically, each neuron is characterized by a critical frequency, f$_c$, typically ranging among neurons between 2 and 30 Hz. Stimulation frequencies above f$_c$ result in NRFs, whereas for supra-threshold stimulations below f$_c$ a response is assured. The probability of the NRFs is such that the neuron functions similar to a low pass filter, saturating its firing rate (**Figure 1**, lower panel). Quantitatively, for a stimulation frequency f, the NRF probability is 1-f$_c$/f, i.e. the firing frequency is saturated at f$_c$. Thus, changing the stimulation frequency will change the probability for NRFs, while the firing frequency remains bounded, f$_c$.

This observation is demonstrated using a cultured cortical neuron, functionally separated from its network by synaptic blockers, with above-threshold stimulations (see section 2.1 *In-vitro* experiments). We examine a neuron with periodic stimulation trials of 10, 12 and 15 Hz, and NRFs appear after a short transient where the neuron exhibits an increase of its NRL (**Figure 1**, upper panel). Examining the temporal firing rate of the neuron (**Figure 1**, lower panel), it is noticeable that the firing frequency of the neuron is saturated at $f_c$=5 Hz, independent of the stimulation frequency.

The effect of NRFs is first examined on small neuronal circuits using the following experiment: We stimulate the neuron ones and impose on the neuron two self-feedback delay loops, e.g. 600 ms and 630 ms (**Figure 2A**). The neuron is stimulated 600 ms and 630 ms after each evoked spike (see section 2.1.6 *In-vitro* experiment with feedback loops and neural circuits). In the case of vanishing probability for NRF, the neuron should fire every 30 ms (Kanter et al., 2011;Vardi et al., 2012b), i.e. 33.3 Hz. Since $f_c$=5 Hz, some NRFs appear (**Figure 2B**) forming bunched firing, separated by ~600 ms, whereas the intra-bunch time-lags, 30 ms, originated from the difference between the two feedback delay loops (**Figure 2B**). The emergence of such firing bunches indicate some dynamical changes of the NRF's probability, where the neuron adapts its failure probability as a result of its recent stimulation history.

A more biologically realistic scenario is a neuronal circuit consisting of three artificially connected neurons (Vardi et al., 2012b), forming the same delay loops (**Figure 2C**), 600 ms and 630 ms. In addition, each neuron is identified by different $f_c$ and only the central neuron receives the initial stimulation. For the central neuron, which is characterized by $f_c$=7.2 Hz, NRFs occur, since the frequency of its driven neurons is greater than its critical frequency, 6.8+4>7.2 Hz, and similarly the outer neurons have response failures (7.2>6.8, 7.2>4) (**Figure 2D**). Besides the formation of firing bunches for each neuron, their firing are correlated at zero- or shifted- time-lags (**Figure 2D**). The question whether the repeated bunches in such small neuronal circuits shed light on macroscopic cortical oscillations requires to sail towards large scale simulations.

## 3.2 Short-term memory of neuronal plasticity

The observed firing bunches indicate a form of short-term memory of neuronal plasticity, where the NRF probability is mainly a function of the few preceding stimulations. Our next goal is to experimentally quantify this neuronal plasticity and then examine its implementation on the dynamics of large neural networks using large scale simulations.

We first assume a simplified network where each neuron has two above-threshold inputs, two outputs and the same $f_c$ (**Figure 3A**), hence the statistics of the inter-stimulation-intervals of each neuron is expected to approximately follow an exponential distribution with $2f_c$ rate, Exp($2f_c$) (**Figure 3B**, upper panel). To quantify the statistics of the NRFs, a long stimulation trail obeying Exp($2f_c$), under $\tau_c/8$ time resolution, was given to a cultured neuron with $f_c$~5 Hz (**Figure 3B**). Next, the conditional probabilities for a successful response (an evoked spike), p(s|$\tau_i,\tau_{i-1}$), were estimated for events where the current inter-stimulation-interval equals $\tau_i$ and the previous one equals $\tau_{i-1}$ (**Figures 3B,C**). It is evident that p(s|$\tau_i,\tau_{i-1}$) is primarily a function of $\tau_i$, i.e. the probability for a successful response is dictated mainly by the current inter-

stimulation-interval, $\tau_i$. Hence, the NRFs can be dynamically approximated using p(s|$\tau$) (**Figure 3D**), indicating that the neuronal response failure probability might be fairly estimated based solely on the last inter-stimulation-interval, $\tau$.

### 3.3 Neuronal oscillations on the network level

The experimental estimation of p(s|$\tau$) is utilized to simulate a large scale network (**Figure 3A**) and is exemplified for 2000 neurons where delays between connected neurons, D, are randomly selected from a uniform distribution $U(10,15)$ ms. The simulation is initialized with timings of evoked spikes for a subset of the neurons, however, besides the transient time results were found to be insensitive to the initial conditions. The response failure is then randomly selected following the experimentally measured p(s|$\tau$), independently for each neuron and stimulation (**Figure 3D**). Indeed, the assumption Exp(2fc) (Figure 3B, upper panel) was confirmed (**Figure 3E**), the statistics of the inter-stimulation-intervals of each neuron approximately follows an exponential distribution with 2$f_c$ rate. The raster plot of the network firing as well as the time-dependent firing rate (**Figures 4A,B**) clearly indicate cooperative oscillation which can be quantified using the Fourier analysis to $f_{osc}$~3 Hz (**Figure 4C**, see section 2.2), and are absent in the Fourier analysis of the firing of each individual neuron (**Figure 3F**). Another broadened Fourier peak is centered at $f_\gamma$~80 Hz, gamma oscillations(Brunel and Wang, 2003;Cunningham et al., 2004;Fries, 2009;Minlebaev et al., 2011;Dugladze et al., 2012), which is attributed to the average delay, <D>=12.5 ms. It reflects the average firing frequency of each neuron where NRFs vanish and all delays are equal to <D>, as GCD=<D> for delay loops of such a random network(Kanter et al., 2011;Vardi et al., 2012a). Similar cooperative oscillations were obtained using a counterpart simulation for the same network (**Figures 4D,F**) while using the theoretical form p(s|$\tau$) (**Figure 3D**). Although the form of p(s|$\tau$) varies among neurons as well as between the theoretical and the experimental forms (**Figure 3D**), the cooperative oscillations are found quantitatively to be only slightly affected by its exact form. The robustness of $f_{osc}$ was also confirmed in simulations for more realistic scenarios where $f_c$ significantly varies among neurons as well as their input and output connectivity distributions. A more biological realization is exemplified in **Figures 4G-I**. The number of connections per neuron is much greater than 2, i.e. more than 50 pre- and 50 post- synaptic connections, where most of them are sub-threshold and on the average 1.5 of pre- and post-synaptic are above-threshold. Specifically, each sub-threshold connection produces an excitatory postsynaptic potential, EPSP, which is equal to 0.03 of the threshold. It is apparent that these additional connections do not qualitatively change the oscillations. In addition, $f_\gamma$ was found to be robust to a wider distribution of delays and followed its center (**Figure 4I**). Note that without these intrinsic NRFs, i.e. p(s|$\tau$)=1, the firing of each neuron is only bounded by the refractory period which is in the order of several milliseconds. In this limiting

case, the abovementioned theta and gamma oscillations disappear, as was confirmed in simulations (not shown).

### 3.4 Cortical oscillations versus statistical properties of the network

The origin of the fast oscillations, $f_\gamma$, (**Figures 4C,F,I**) is evident, however, the mechanism underling the slow cooperative oscillations(Wu et al., 1999;Sanchez-Vives and McCormick, 2000;Bollimunta et al., 2008;Nir et al., 2008;Crunelli and Hughes, 2010;Bollimunta et al., 2011), $f_{osc}$, is still unclear. To explore this mechanism we identify the following two characteristic distances on the network. The first distance, Path, is the average minimal path between pairs of nodes, and the second distance, Loop, is the average over the minimal feedback loop of each node, both counted by the number of nodes along the route (**Figure 5A**). Numerical estimations based on various network topologies indicate that the average values of these two distances as well as their distributions are almost identical (**Figure 5B**) and their scaling decrease as $1/\ln(<K>)$, where $<K>$ is the average neuronal input connectivity (**Figure 5C**). These identities and scaling (**Figures 5B,C**), are also supported by the following theoretical argument. Assume a random network consisting of N neurons and an average connectivity $<K>$. The quantity Q(m) denotes the number of new connected neurons to a seed neuron at a distance of m neurons, hence Q is proportional to the probability (green) in **Figure 5B**. Start at an arbitrary neuron, Q(0)=1, this neuron is connected (pre-synaptic) to Q(1) neurons. Using a recursive formula one can approximate

$$Q(i) = N \left( 1 - \exp\left( -<K> \frac{Q(i-1)}{N} \right) \right) \left( 1 - \sum_{m=1}^{i-1} \frac{Q(m)}{N} \right)$$

where N(1-exp(-<K>Q(i-1)/N)) stands for the average number of neurons at a distance i, which are connected from new neurons at distance i-1. The rightmost term is the probability that the neuron at distance i is a new neuron which was not counted at shorter distances, m<i. This recursive relation is solved numerically and the normalized Q(i) are presented in **Figures 5B**. The above analysis is valid for Loops and Paths, hence their statistics are identical, in agreement with the sampling of these quantities in **Figures 5B**.

The importance of this distance in the formation of $f_{osc}$ can be understood according to the following argument. Assume a random subgroup of firing neurons activates another random group of neurons and vice versa. The minimal delay between pairs of neurons belonging to the two subgroups is Path·<D>. Consequently, the oscillations are expected to scale with Path·<D>, and indeed results indicate that $f_{osc} \propto \ln(<K>) \propto (Path)^{-1}$ (**Figures 5C,D**). The minimal path is the most reliable one with respect to the NRFs, however, the effect of longer paths is not negligible as they might maintain the activity of the Path (**Figure 2**), especially as their entropy is higher. In addition, the NRFs are responsible to limit the firing of all the network simultaneously (**Figures 4B,E,H**). Assume one neuron fires and activates <K> neurons after <D> ms, hence after m<D> ms, <K>$^m$ neurons fire. This exponential firing growth is bound by m≈Path, as self-feedback loops (**Figures 2A,B**) significantly lead to NRFs and to a fast decrease in the firing rate. In addition, for a given network topology, $f_{osc}$ is found to scale with ln($f_c$) (**Figure 5E**), and to be robust for networks composed of neurons with different $f_c$ (**Figures 4G,H,I**). These predictions might be realized in further experiments by controlling the network topology either by the neuronal concentration or by pharmacological manipulations.

### 3.5 Analytical description of the network oscillations

An analytical description of $f_{osc}$ is also possible, and to simplify the presentation the method is briefly described for homogeneous networks only. Each node has the same fixed input and output connectivity, *K*, and all delays are equal to *D* ms, hence neurons can fire only at *i·D* ms, where *i* is an integer. The fraction, *R(m)*, of neurons that fire at step *m* is given by

$$R(m) = \chi(m) \cdot K \cdot R(m-1) + \xi(m) \qquad (1)$$

where *ξ(m)* stands for the time-dependent white noise representing the stochastic nature of the response failures. The function *χ(m)* represents the susceptibility of the network, i.e. the fraction of neurons that fires if all neurons are stimulated at step *m*, and is explicitly given by

$$\chi(m) = \sum_{i=0}^{\infty} p(s|i) \cdot K \cdot R(m-(i+1)) \prod_{n=2}^{i} (1 - K \cdot R(m-n)) \quad . \quad (2)$$

The first term, p(s|i), stands for the probability for an evoked spike when the previous stimulation was given before *i* steps (**Figure 3D**). The term *K·R(m-(i+1))* pinpoints a neuron stimulated before *i* steps, and the product, the rightmost term, indicates that the neuron was not stimulated since step *(m-i)*. Equation (2) indicates that *χ(m)* is a function of the variable *R(l)* only, with *l<m*. Hence, after the insertion of equation (2), into equation (1), one finds a recursion relation for *R(m)* which can be solved numerically given the initial conditions. The dynamical solution of this recursion relation revealed oscillations which were found to fit fairly good with those observed in large-scale simulations (**Figures 4C,F,I**). The equations imply that the network has some memory of its previous activity, which dictates the responsiveness of the entire network. This analytical description can be generalized to advanced structured networks, including random connectivity, distribution for the delays as well as to include variations among neuronal critical frequencies, $f_c$.

## 4. Discussion

We have demonstrated that intrinsic neuronal response failures drive a neural network activity towards oscillations, where high frequency oscillations, gamma, and low frequency oscillations, delta and theta, coexist. The high frequency oscillations correspond to the average delay between connected neurons in the network, while low frequency oscillations are governed by statistical properties of the network, e.g. the average number of connections per neuron and the average critical frequency of neurons. The coexistence of high and low frequency oscillations was confirmed in a new type of simulations, based on a single neuron *in-vitro* experiments, to evaluate the firing activity of complex networks. Results were also supported by an analytical description of the stochastic dynamics of the network.

Preliminary results *in-vivo* support our findings. The NRL increases by several milliseconds under periodic stimulations and terminates at an intermittent phase(Vardi et al., 2015). This phase is characterized by fluctuations around a constant NRL and accompanied by NRFs. Results indicate that $f_c$ can be below 10 Hz and vary among neurons. However, quantitative measurements of $f_c$ and the statistics of the NRFs require long stimulation trials, i.e. many thousands of high frequency periodic stimulations, which are currently beyond our experimental capabilities.

The average firing rate of neurons in the network is low, e.g. ~3.6, ~3.9 and ~2.6 Hz in **Figures 4B,E,H**, respectively. These network low firing rates are lower than the neuronal critical frequency in **Figures 4B,E**, $f_c$=5.7 Hz, and $<f_c>$=6.5 Hz, in **Figure 4H**. Surprisingly, the neuronal critical firing frequency is not saturated even when the network is completely excitatory. A biological mechanism that suppresses the firing frequency of a single neuron below $f_c$ is aperiodic time-lags between stimulations (Vardi et al., 2015). For illustration, assume a slow mode of alternation between stimulation frequencies of $2f_c$ ($0.5\tau_c$) and $2f_c/3$ ($1.5\tau_c$), such that $<\tau>=\tau_c$. For the high and low frequency mode, the expected probability for a NRF is 0.5 and 0, respectively. Consequently, $<ISI>$=$0.5(1.5\tau_c+\tau_c)$=$1.25\tau_c$, corresponding to a lower firing rate, $0.8f_C$. In addition, the firing rates are considerably lower than $f_\gamma$, indicating that high frequency network oscillations consist of temporarily synchronized sub-groups of neurons. Indeed, the Fourier spectrum of a single neuron does not exhibit any dominant peaks (**Figure 3F**).

Although the formation of broadband network oscillations is usually attributed to the existence of inhibition, it is evident that another possible mechanism is intrinsic neuronal response failures that dynamically drives neural networks to generate coherent oscillations with low averaged firing rates(Vardi et al., 2015). These observations raise the question of which functionalities demand synaptic inhibition. It was shown that inhibition slightly suppresses the network firing frequency even further(Vardi et al., 2015) and it also might change the amplitude of the oscillations. An additional possible hypothesis is that the role of inhibitory connections is to allow some neuronal computations which are based on conditional temporal formation of neuronal firing patterns. This type of functionality is an exclusive property of inhibitory synapses which probabilistically block an evoked spike of its driven nodes in a given time window(Vardi et al., 2013b;Goldental et al., 2014). In addition, the coexistence of the network oscillations with neuronal inhibition is intriguing, and especially the questions whether inhibition induces more modes of oscillations, sharpens the existing ones, or suppresses the oscillatory behavior and stabilizes the network activity. Preliminary results of simulations indicate that inhibition might suppress the amplitude of the oscillations in the low frequency range and sharpen the oscillations in the gamma range. However, results might be sensitive to the selected parameters.

The interplay between the presented spontaneous cortical oscillations and external stimulations given to the network is another intriguing question. Specifically, it is interesting to examine the coexistence and the interplay between the spontaneous oscillation frequencies determined by the network topology and the frequencies of the induced external stimulations. The understanding of these dynamics will shed light on the emerging cortical oscillations among coupled networks characterized by different statistical properties.

## ACKNOWLEDGEMENTS

This research was supported by the Ministry of Science and Technology, Israel. The authors declare no competing financial interests.

## AUTHOR CONTRIBUTION

R.V. and S.S. prepared and performed the experiments; R.V., S.S. and A.G. analysed the data; P.S. designed the experimental real-time interface; A.G. performed the simulations and developed the theoretical framework with help of P.S.; I.K., R.V. and A.G. wrote the manuscript.

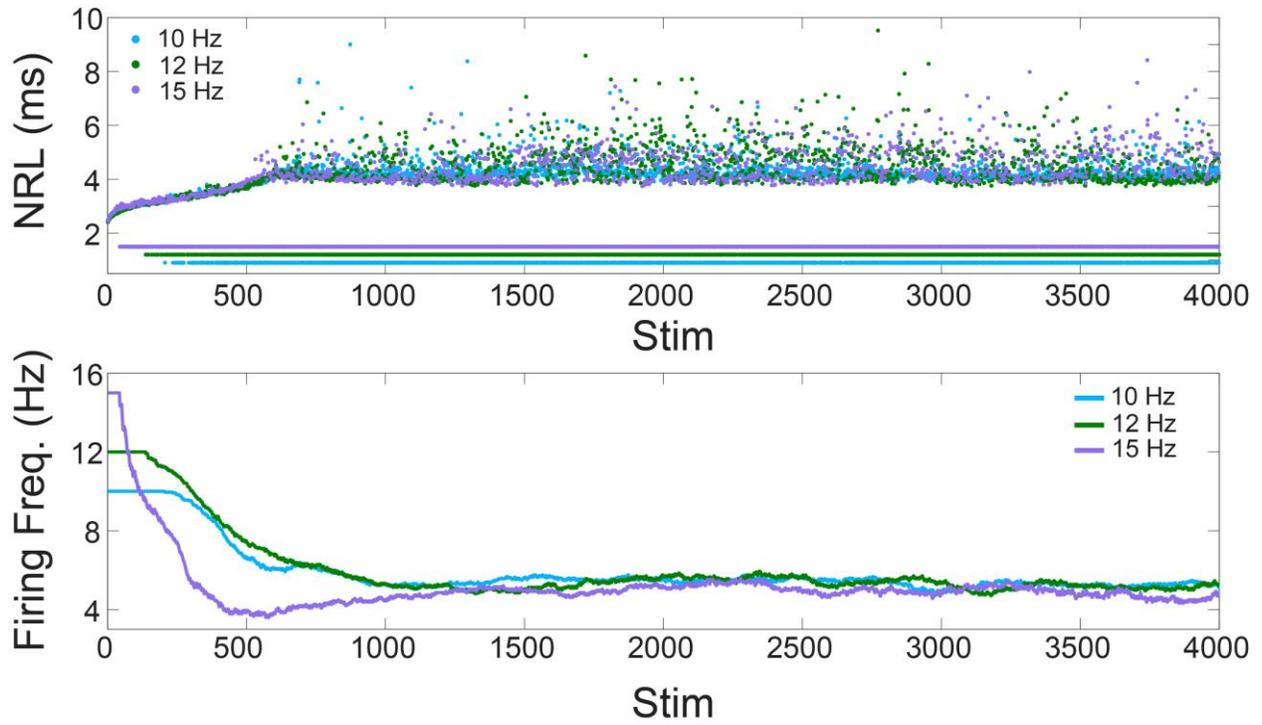

**FIGURE 1 | Neuronal plasticity – *in-vitro* experiments.** Upper panel: The NRL of a neuron stimulated at 10, 12 and 15 Hz (light blue, green and purple dots, respectively). Response failures are denoted by NRL<2 ms. Lower panel: The firing frequency calculated from the averaged ISI using sliding windows of 250 stimulations, or the maximal available one for Stim<250.

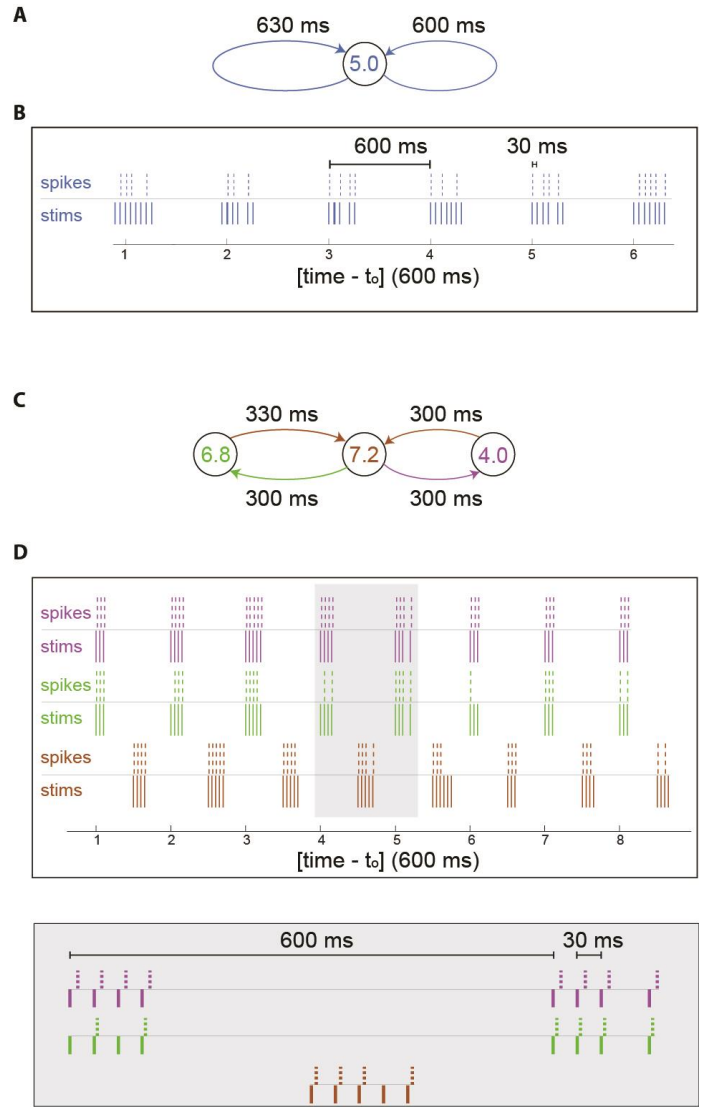

**FIGURE 2 | Firing bunches stem from neuronal response failures – *in-vitro* experiments. (A)** Schematic of the neuron in **Figure 1** characterized by $f_c$=5 Hz, with 600 and 630 ms self-feedback loops. **(B)** A 3.5 s snapshot of the experimental results of **(A)** where stimulations (blue lines) and their corresponding evoked spikes (blue dashed lines) were recorded after an offset of $t_o$=21 s and a preparation at 5 Hz stimulation frequency over 300 s to settle the neuron at the intermittent phase. **(C)** Schematic of a circuit consisting of three different neurons with $f_c$=6.8 (green), 7.2 (brown) and 4.0 (purple) Hz and 600 and 630 ms delay loops, similar to **(A)**, (different neurons than the one in **(A)**). **(D)** A 5 s snapshot of the experimental results of **(C)** where ten initial stimulations were given to the central neuron (brown) at 4 Hz and $t_o$=25 s. Stimulations given to the colored neurons (brown/green/purple lines, respectively) and their corresponding evoked spikes (brown/green/purple dashed lines, respectively). A zoom-in of the gray area is presented (bottom).

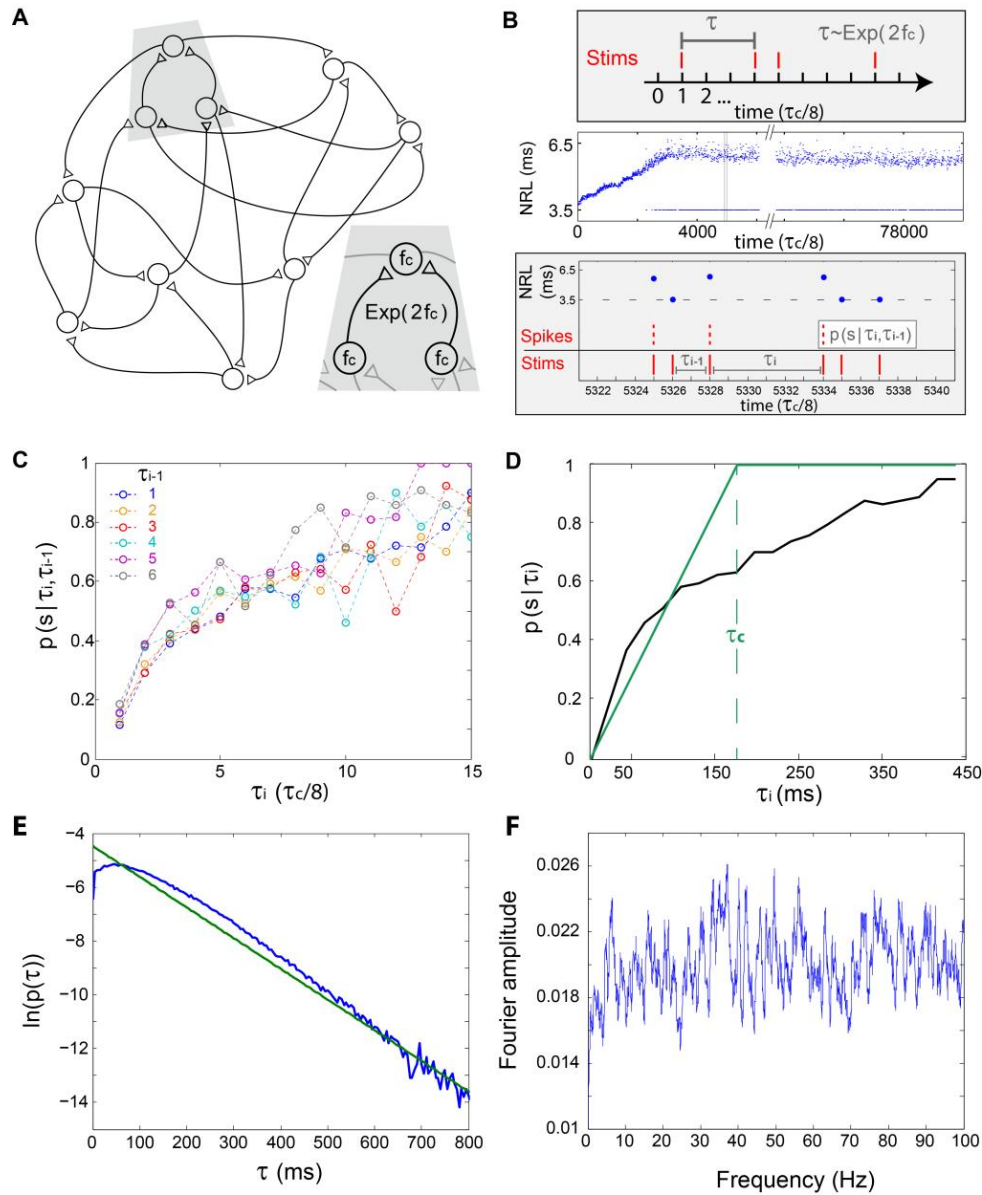

**FIGURE 3 | The short-term memory of the stochastic neuronal response failures – *in-vitro* experiments and simulations. (A)** Schematic of a prototypical examined excitatory network, where each neuron has two pre- and two post- synaptic connections and the same $f_c$. Inset: The time-lags between neuronal stimulations is expected to approximately follow an exponential distribution, Exp($2f_c$). **(B)** Upper panel: The stimulation scheme where a neuron is stimulated under the resolution of $\tau_c/8$, such that the discrete differences between two consecutive stimulations, $\tau$, follows Exp($2f_c$). Middle panel: Experimental NRL of a cultured neuron with $f_c \sim 5$ Hz under a long trial of stimulations following Exp($2f_c$), response failures are denoted at NRL=3.5 ms. Lower panel: Zoom-in of the middle panel (gray area) and schematic of the conditional probability p(s|$\tau_i$,$\tau_{i-1}$), measuring the probability of a successful response, spike, given that the current inter-stimulation-interval equals $\tau_i$ and the previous one equals $\tau_{i-1}$. **(C)** The probabilities

$p(s|\tau_i,\tau_{i-1})$ obtained from the experimental data in **(B)** for time>3500 (middle panel). Each colored line presents $p(s|\tau_i,\tau_{i-1})$ for a given $\tau_{i-1}$ in $\tau_c/8$ time units (legend). **(D)** The experimentally measured $p(s|\tau_i)$ (black), measuring the probability of a successful response, spike, given that the current inter-stimulation-interval equals $\tau_i$, and the theoretically predicted one (green) using the simplified assumption, $p(s|\tau_i)=\tau_i/\tau_c$ for $\tau_i<\tau_c$. For both curves, the average ISI~$\tau_c$ is preserved. **(E)** The probability density function of inter-stimulation-intervals, $\tau$, for all neurons of **Figure 4A** (blue). **(F)** Typical Fourier amplitude of spike timings of a randomly chosen neuron, taken from **Figure 4A**.

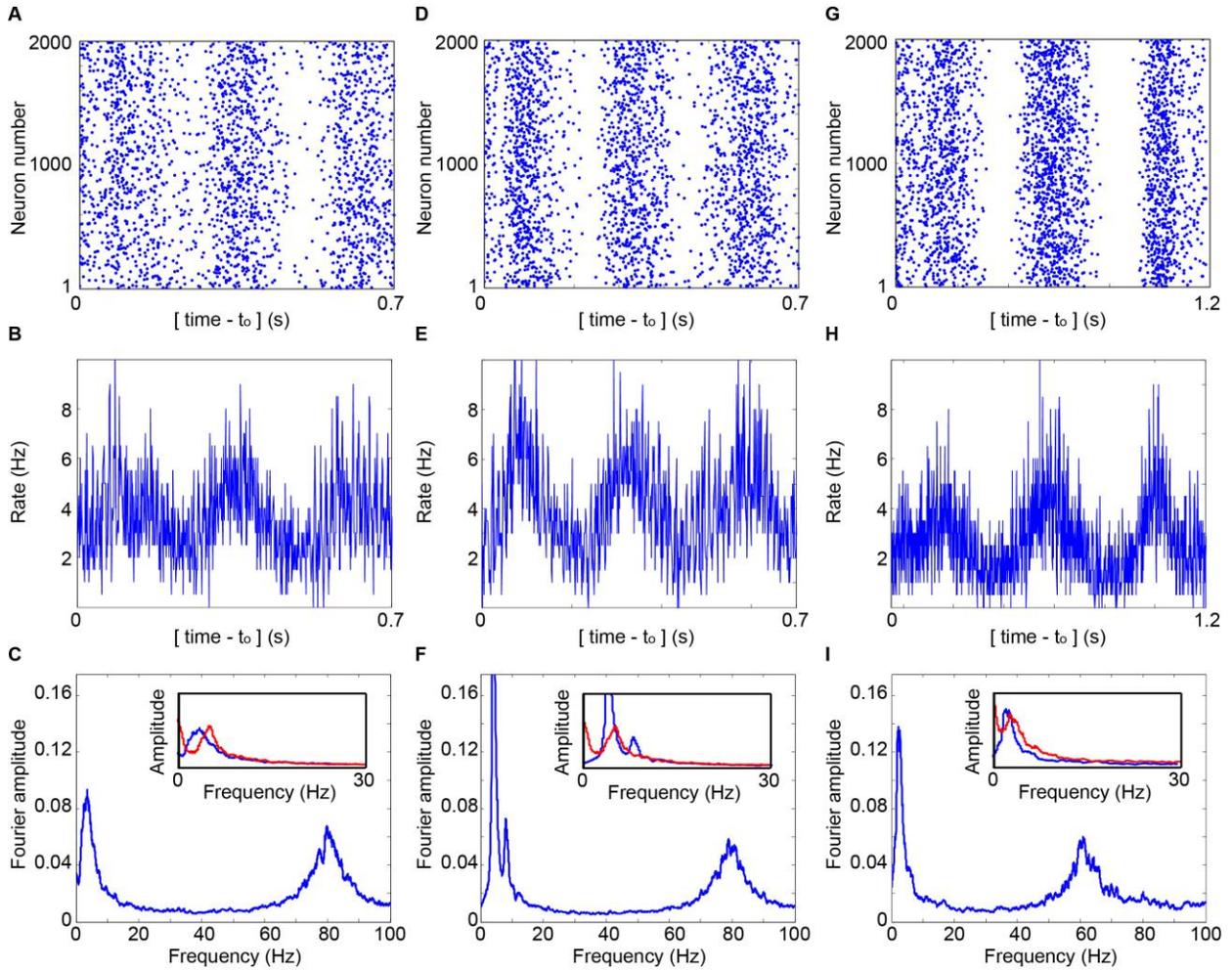

**FIGURE 4 | Cooperative cortical oscillations on a network level. (A)** Raster plot of the evoked spikes (blue dots) obtained in the simulation of a network of 2000 neurons with $f_c$=5.7 Hz. Each neuron has two randomly selected pre- and post- synaptic connections, and the simulation is based on the experimentally obtained p(s|τ), (**Figure 3D**). Delays are randomly selected from $U(10,15)$ ms. The contrast of the raster was enhanced using a dilution of a constant amount of randomly chosen points in each sliding window of 23 ms, with a step of 0.23 ms. The average dilution is ~60% of the points. **(B)** The average firing rate per neuron as a function of time, calculated for windows of 1 ms. **(C)** The normalized Fourier amplitude, using a sliding window of 1 Hz, of the entire firing of all neurons over a time slot of 30 s, indicating $f_{osc}$~3.6 Hz and $f_\gamma$~80 Hz. Inset: The normalized Fourier amplitudes in the range [0,30] Hz obtained from R(m), equation (1), D=12.5 ms (red) and from the simulation, ((**B),** blue). **(D-F)** Similar to **(A-C)** where each neuron has on average two pre- and post- randomly chosen synaptic connections and utilizing the theoretical p(s|τ), (**Figure 3D**). $f_{osc}$~4.0 Hz and $f_\gamma$~80 Hz at **(F)**. **(G-I)** Simulation of a network of 2000 neurons where each neuron has on the average 1.5 pre- and 1.5 post- synaptic above-threshold connections, and 50 pre- and 50 post- synaptic sub-threshold connections with a strength of 0.03, relative to a threshold of 1. p(s|τ$_i$)

is generalized to an exponential decay function of the neuronal stimulation history, $\Sigma(\tau_{i-m}/\tau_c)\exp(-\alpha m)/[\Sigma\exp(-\alpha m)]$, $\alpha=1.4$ and the sum is over stimulation history, $m\geq0$. Delays are randomly selected from $U(12.5,20)$ ms and $f_c$ from $U(3,10)$ Hz. $f_{osc}\sim2.6$ Hz and $f_\gamma\sim60$ Hz at (**I**), and the inset is similar to (**C**) and (**F**), but with K=1.5, $f_c=6.5$ Hz and D=16.25 ms.

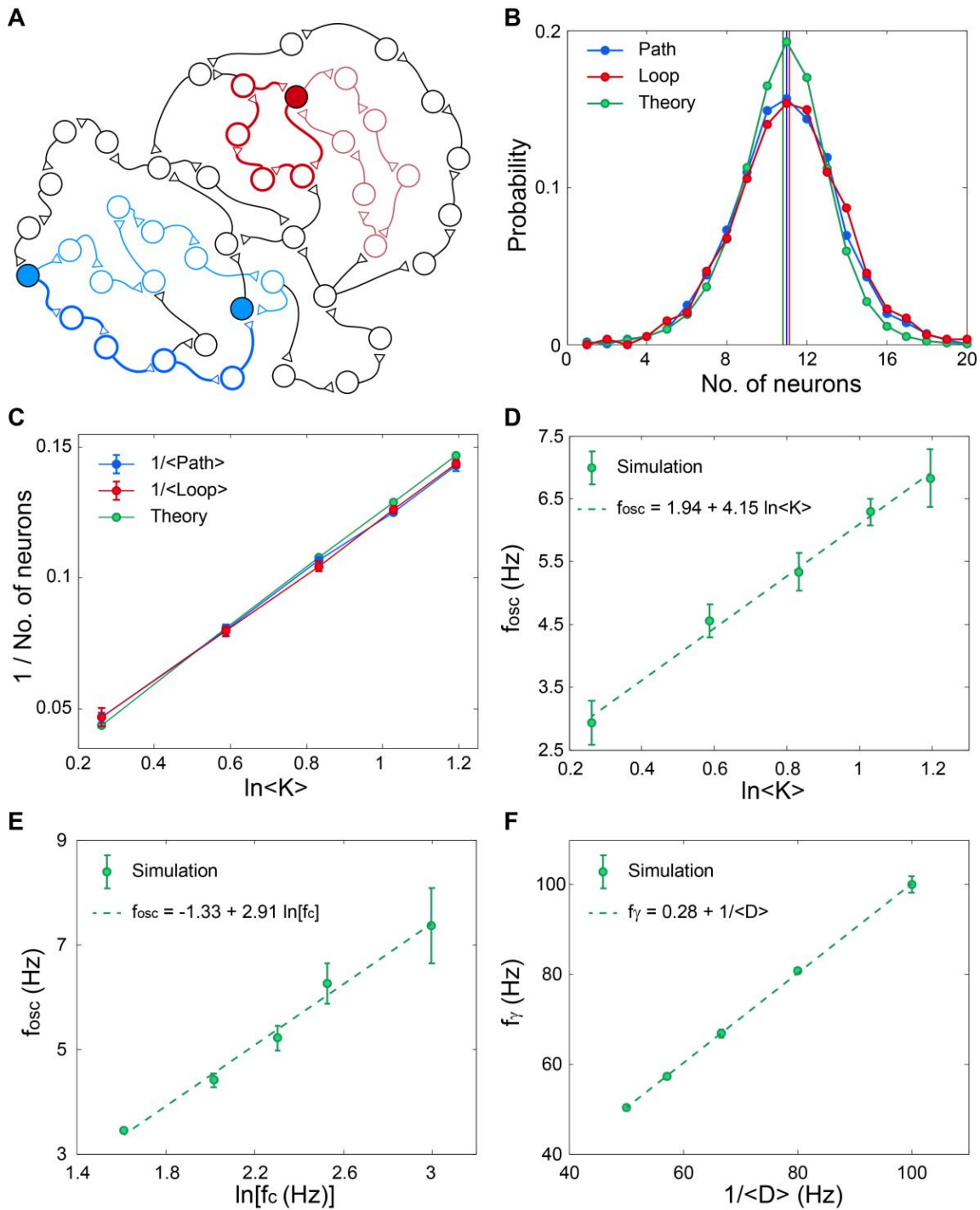

**FIGURE 5 | Scaling properties of $f_{osc}$ and $f_\gamma$. (A)** Schematic of an excitatory network where a self-feedback loop (light-red line) and the minimal self-feedback loop (red line) for a given neuron (filled red circle) are denoted. Similarly, a path (light-blue line) between two neurons (filled blue circles) and the minimal path (blue line) are denoted. **(B)** The distribution and its average (vertical lines) for the minimal path (Path) and for the minimal loop (Loop) obtained in simulations for networks as in (**Figure 3A**) with

N=4000, error bars are comparable with the circles. The analytical estimation is shown in green. **(C)** The scaling of the averaged quantities in **(B)** as a function of the average connectivity, <K>. **(D)** Simulation results indicate $f_{osc} \propto \ln$<K>, where N=4000, $f_c$ is randomly chosen for each neuron from $U$(5,15) Hz and delays are randomly chosen from $U$(10,15) ms. The probability for a connection between two neurons is <K>/N. Error bars indicate the standard deviation. **(E)** Simulation results indicate $f_{osc} \propto \ln(f_c)$, for networks as in **(D)**, with <K>=2, but $f_c$ is the same for all neurons. **(F)** Simulation results indicate $f_\gamma \propto 1/$<D>, for networks as in **(D)**, with <K>=2, but delays are randomly chosen from $U$(<D>-2.5,<D>+2.5) ms.